\documentclass[]{aa}
\usepackage{color}
\usepackage[table]{xcolor}
\usepackage{colortbl}

\newcommand{\avg}[1]{\ensuremath{\langle #1 \rangle}}

\begin{document}
\title{Simulations of facular magnetic fields on cool stars}
\subtitle{I: Main-sequence stars with solar metallicity}
\author{Tanayveer Singh Bhatia\inst{1}\fnmsep\thanks{Corresponding author: bhatia@mps.mpg.de}
        \and Robert H. Cameron\inst{1}
        \and Sami K. Solanki\inst{1}
        \and Damien F. Przybylski\inst{1}
        \and Veronika Witzke\inst{1,2}
        \and Alexander Shapiro\inst{1,2}
        \and Nadiia Kostogryz\inst{1}
        }
\institute{\inst{1}Max Planck Institute for Solar System Research, Justus-von-Liebig Weg 3, D-37077, Göttingen, Germany\\
\inst{2}Institute of Physics, University of Graz, Universitätsplatz 5, 8010 Graz, Austria\\}
\date{Received XXXX}

\abstract
{
{\textit{Context:} Stellar convection in the presence of magnetic field affects the emergent intensity, as well as the structure and evolution of cool main-sequence dwarfs.}\\
{\textit{Aim:} We aim to understand the effect of faculae-like field strengths on near-surface stellar convection using 3D radiative MHD simulations of near-surface magneto-convection.}\\
{\textit{Methods:} We compare simulations of F, G, K, and M main-sequence stars with a small-scale dynamo (SSD) to faculae-like spatially averaged field strengths (from 100 to 500 G). We focus on the effect of imposed magnetic field on the thermodynamic stratification and velocities, along with the bolometric intensity and surface field strength}.\\
{\textit{Results:} Imposed magnetic fields result in reduced average density and gas pressure near the surface compared to the SSD simulations. The temperature stratification also shows a dip at and just below the stellar surface. The changes in average bolometric intensity are within a percent, with different trends with field strength for different stellar types. In addition, the convective velocities are reduced. The magnitude of changes in thermodynamic quantities are related to field strength as well as the stellar $T_{\rm eff}$.}\\
{\textit{Conclusion:} Faculae-strength magnetic fields modify the near surface convection by reducing gas pressure and density as well as suppressing convection in regions with strong field concentrations. The strength of these effects depends on the stellar type.}
}


\maketitle

\section{Introduction} \label{sec:intro}

Magnetism is ubiquitous in outer convection zones of cool stars. Evidence of magnetism appears not just in indicators of chromospheric activity \citep{Baliunas1985}, stellar flares \citep{valera2024} and photometry \citep{Strassmeier2009}, but also in polarization \citep{zdi_review}. From an observational point of view, magnetic fields can have a wavelength-dependent signature in the emergent stellar spectrum and limb darkening \citep{ludwigCLV2023,kostogryz2024}, which affects the interpretation of stellar light curves. This is especially important when it comes to characterizing exoplanets and their atmospheres around magnetically active stars \citep{deWit2013}. Accounting for stellar magnetism is of prime importance in the field of exoplanet detection and characterization. In the post-\textit{Kepler} era, the detections of exoplanets from photometry has shot up to thousands of candidates. These candidates are then targeted by ground-based radial velocity (RV) campaigns to accurately characterize the properties of the exoplanet candidate. The biggest obstacle to precise and accurate characterization is stellar variability \citep{eprv2021}. In addition, transmission spectroscopy, which is used to characterize exoplanetary atmospheres and the wavelength-dependent radius of exoplanets, is also affected by stellar activity \citep{rackham2023}. Hence, there is a requirement for models that capture the radiative MHD properties of faculae and spots as accurately as possible.

The most widely used models for interpreting stellar spectra are simplified, hydrodynamic, 1D models that either do not account for magnetism at all \citep{atlas,phoenix,marcs} or implement only the surface radiative effects of starspots (i.e., flux blocking) in a parameterized manner \citep{somers2020}. A more consistent approach to modeling magnetic fields in a mixing length framework was carried out by \citet{feiden2013}.

Early studies of non-radiative compressible 3D MHD convection have focused on the types of convective patterns that form over a range of imposed vertical field strengths and the transitions between these patterns with varying field strength. Chapter 8 of \citet{WeissProctor2014} gives an overview of these studies in a dimensionless setup. To summarize, it was found that the convective pattern depends strongly on the strength of the imposed vertical magnetic field. The magnetic structures ranged from point- and ribbon-like concentrations at weaker field strengths to large-scale flux-separated weak-field convecting regions and strong-field regions with suppressed convection \citep{WeissProctor2002} at stronger field strengths. In the presence of the strongest fields, regular convection gets replaced by over-stable small-scale oscillations.

In the last four decades or so, the sophistication of numerical models has increased in tandem with available computing resources, allowing realistic 3D radiative MHD simulations of near-surface convection with magnetic fields. Recently, simulations of faculae-like magnetic fields \citep{beeck3,salhab2018} as well as spots \citep{panja2020,starspots2025} on cool main-sequence stars have been conducted with a box-in-a-star approach to study the effect of magnetism on structure as well as the wavelength dependence of intensity \citep{norris2023,smitha2025,witzke2022}.

The present work is an update as well as an extension to the facular studies of \citet{beeck3}. We focus more on the effects of field on convection and structure in this paper, in addition to the effects on stellar photospheric velocities, intensity, and field distribution as in \citet{beeck3}. To our knowledge, no other study has probed this aspect for realistic stellar surface simulations, except for the analysis in \citet{paper3}, which was restricted to SSD simulations. In addition, our reference cubes are saturated SSD atmospheres already containing a basal level of magnetism, instead of being purely hydrodynamic, and the simulations are run with a new equation of state \citep{freeeos}. We note that both the equation of state as well as the opacities for radiative transport use abundances from \citet{asplund2009}, which is not the case for \citet{beeck3}.

The paper is organized as follows: in Sec. \ref{sec:methods}, we describe the setup as well as the data analysis choices, in Sec. \ref{sec:res}, we describe the changes in thermodynamic and convective structure, and in Sec. \ref{sec:disc}, we provide explanations for these changes and their implications for our understanding of magnetized near-surface stellar convection. Finally, in Sec. \ref{sec:conc} we summarize the main results.

\section{Methods}\label{sec:methods}

In this paper, we considered the stellar types F3V, G2V, K4V, and M0V and conducted near-surface simulations using a box-in-a-star approach. We used the \texttt{MURaM} radiative MHD code \citep{muram1,muram2} to conduct all the simulations. We employed the small-scale dynamo (SSD) models described in \citet{paper1,paper3} as initial conditions. Briefly, the simulation boxes are periodic in the horizontal directions and have an open (symmetric) bottom boundary for flows and magnetic fields. Two parameters determine the stellar type: the effective temperature, $T_{\rm eff}$, and the surface gravity, $g_{\rm surf}$. The latter was set as a constant in these simulations, whereas the former was determined indirectly by specifying the entropy of inflows at the bottom boundary. In addition, specifying the pressure at the bottom boundary sets the height of the photosphere in the simulation domain. For more details of the models and setup, we refer to the publications cited above.

We then imposed uniform vertical fields of 100 G, 200 G, 300 G, and 500 G on the existing SSD simulations. After inserting the fields, the simulations were run for a few hours up to the point where variations in root-mean-square velocity throughout the box and the effective temperature had stabilized. This allowed the field to organize itself naturally.

Up to this point, the simulations were run with gray opacity (no wavelength dependence). The next step was to introduce a more realistic treatment of radiative transfer. The basic idea is to compute the average opacity for specific “bins” in wavelength and a reference optical depth ($\tau_{\rm ref}$), instead of solving the radiative transport equation for a huge number of wavelength points. For the present set of simulations, 12 opacity bins were used, with thresholds in wavelength and $\tau_{500 \rm nm}$ formation height similar to those used in Table 2 of \citet{beeck2012} for the \texttt{STAGGER} simulations. The bins themselves were averaged from an opacity distribution function \citep{Kurucz1979,atlas2021} following the approach of \citet{Voegler2004}. The choice of 12 bins was made based on the excellent match obtained between solar surface simulations using MURaM and solar observations of the irradiance spectrum, photospheric line strengths, shapes (including bisectors), as well as limb darkening \citep{witzke2024}. A description of the binning procedure is provided in Appendix \ref{app:opac}. For the imposed field simulations, we used the final SSD opacity bins. At this stage, the F-star simulations were doubled in resolution in the vertical direction to better resolve the relatively sharp gradients near the optical surface, since we plan to use these cubes for future spectral synthesis.

\begin{table}[ht]
    \centering
    \caption{Color palette for line plots covering all the simulations.}
    \begin{tabular}{>{\bfseries}l *{5}{ >{\centering\arraybackslash}m{2em}}}
        \hline
        \hline
        & SSD & 100G & 200G & 300G & 500G \\
        \hline
        \textbf{F-star} & 
        \cellcolor[rgb]{0.17914648,0.49287197,0.73542484} & 
        \cellcolor[rgb]{0.29098039,0.5945098,0.78901961} & 
        \cellcolor[rgb]{0.41708574,0.68063053,0.83823145} & 
        \cellcolor[rgb]{0.57960784,0.77019608,0.87372549} & 
        \cellcolor[rgb]{0.71618608,0.83320261,0.91615532} \\
        \hline
        \textbf{G-star} & 
        \cellcolor[rgb]{0.37130334,0.37130334,0.37130334} & 
        \cellcolor[rgb]{0.47843137,0.47843137,0.47843137} & 
        \cellcolor[rgb]{0.58608228,0.58608228,0.58608228} & 
        \cellcolor[rgb]{0.71058824,0.71058824,0.71058824} & 
        \cellcolor[rgb]{0.80878124,0.80878124,0.80878124} \\
        \hline
        \textbf{K-star} & 
        \cellcolor[rgb]{0.18246828,0.59332564,0.30675894} & 
        \cellcolor[rgb]{0.29490196,0.69019608,0.38431373} & 
        \cellcolor[rgb]{0.45176471,0.76708958,0.46120723} & 
        \cellcolor[rgb]{0.59607843,0.8345098,0.57882353} & 
        \cellcolor[rgb]{0.72312188,0.88961169,0.69717801} \\
        \hline
        \textbf{M-star} & 
        \cellcolor[rgb]{0.85033449,0.14686659,0.13633218} & 
        \cellcolor[rgb]{0.94666667,0.26823529,0.19607843} & 
        \cellcolor[rgb]{0.98357555,0.41279508,0.28835063} & 
        \cellcolor[rgb]{0.98745098,0.54117647,0.41568627} & 
        \cellcolor[rgb]{0.98823529,0.67154171,0.56053825} \\
        \hline
    \end{tabular}
    \label{tab:colors}
\end{table}

All simulations were run for 60 minutes after converging to a statistical steady state. Only the output produced during this extra runtime was analyzed. The analysis was conducted over snapshots with a cadence of 90 seconds. We used the same notation as the one described in \citet{paper3} for temporal and horizontal means: overline $\overline{q}$ denotes temporal averages and angular brackets $\avg{q}$ denote spatial averages of the analyzed quantities. The transparent error bars, unless otherwise specified, are $1\sigma$ standard deviation in averages over time. The horizontal averages were computed on geometric slices (unless otherwise stated) and are plotted against the average number of pressure scale heights computed for the SSD case $(\log(p/p_0)$, where $p_0$ is the mean pressure at $z_{\avg{\tau}=1}$). Unless otherwise specified, $\tau$ refers to the Rosseland mean optical depth. We provide a color palette in Table \ref{tab:colors} to show the list of colors used for plots showing multiple simulations.

\section{Results}\label{sec:res}

\begin{figure*}
    \centering
    \includegraphics[width=17cm]{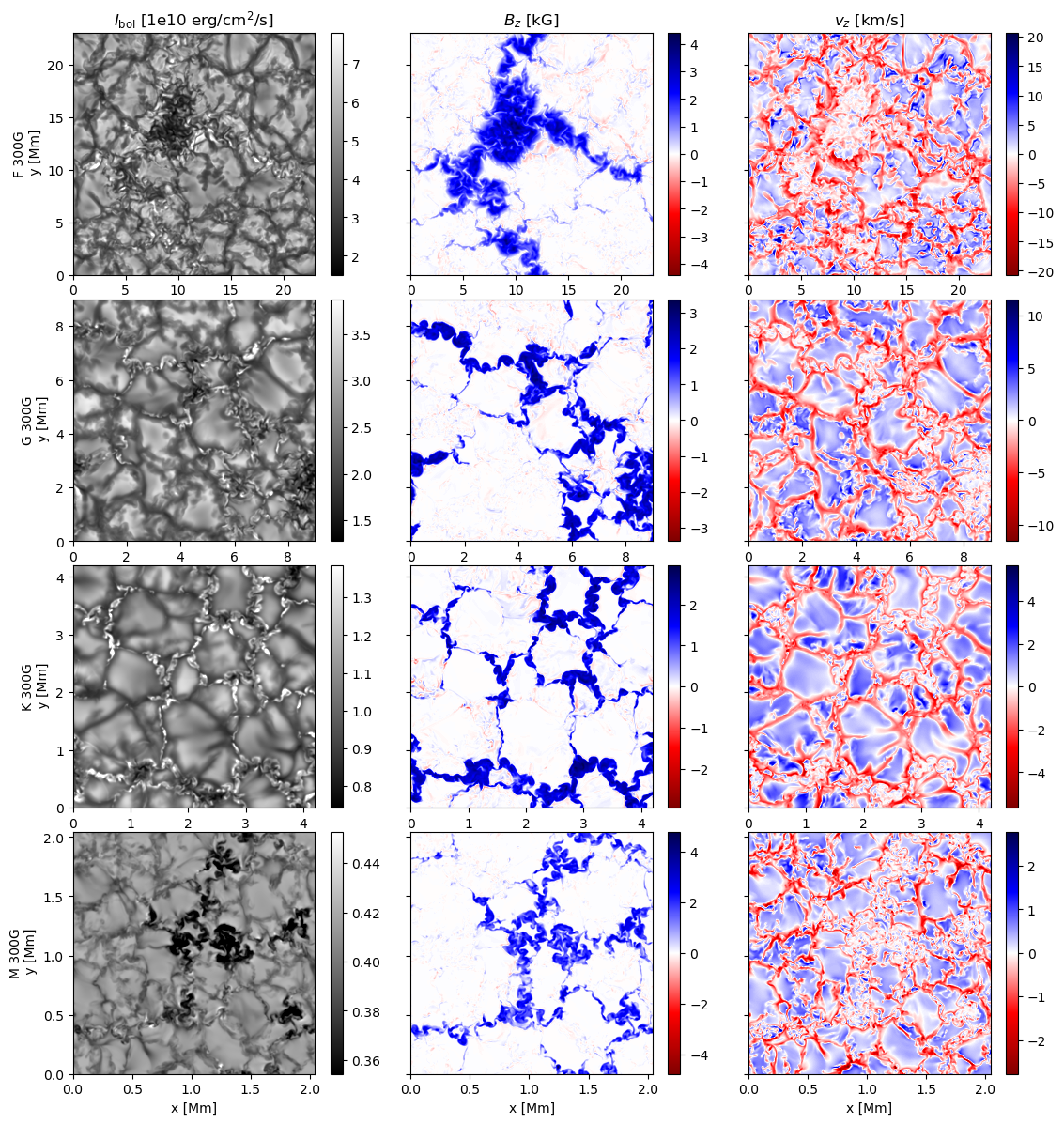}
    \caption{Snapshot of the bolometric intensity $I$ (in $10^{10}$ ergs per centimeter squared per second) for the 300 G case (\textit{left column}) and the corresponding vertical magnetic field, $B_z$ (in kilogauss) (\textit{middle column}), and the vertical velocity, $v_z$ (in kilometers per second) (\textit{right column}), for spectral types (\textit{from top to bottom}) F, G, K, and M, respectively. The field and velocity plots correspond to the $\tau=1$ optical surface.}
    \label{fig:snap_300g}
\end{figure*}

\begin{figure}[ht]
    \resizebox{\hsize}{!}
    {\includegraphics{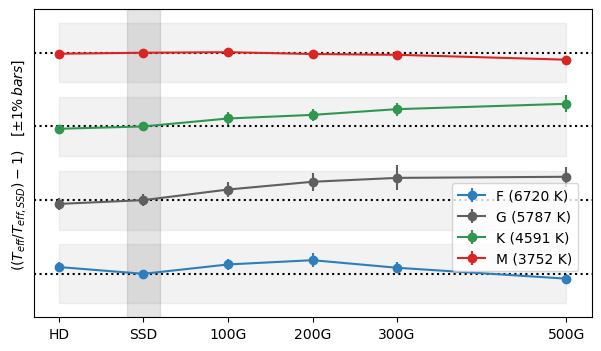}}
    \caption{$T_{\rm eff}$ as a function of field strength, normalized by the SSD $T_{\rm eff}$. The shaded regions represent changes of $\pm 1\%$. The vertical bars at each point are $1\sigma$ average variation in $T_{\rm eff}$. The legend contains the average SSD $T_{\rm eff}$ in Kelvin for each stellar type.}
    \label{fig:teff}
\end{figure}

The effect of magnetic fields on the near-surface structure, convection, and bolometric intensity is varied and depends on stellar type as well as field strength. Stronger fields (500 G simulations) result in the formation of field concentrations that appear dark in intensity (micro-pores). This is due to the stabilizing effects of strong vertical magnetic field on convection \citep[Eq. 1.1]{GT1966}. Weaker fields (300 G and less) result in the formation of bright points and fibrils for the F, G, and K stars. Magnetic fields evacuate intergranular lanes, reducing the density and, accordingly, the opacity, and allowing radiation to escape from surrounding hotter upflows in deeper layers (also termed the “hot-wall” effect) \citep{Spruit1976_bpts}.

Cooler stars (M0V) have higher surface densities and, for the same field strength, are more likely to form darker concentrations. Cooler stars have a lower contrast between upflows and downflows in density and temperature \citep[Fig. 12]{beeck1}. Since the corresponding optical depression in the $\tau=1$ layer in downflows is also reduced, the hot-wall effect is not so effective. These results are quite consistent with previous studies of faculae-like simulations of near-surface convection in cool dwarfs \citep{beeck3,salhab2018}. Fig. \ref{fig:snap_300g} shows a snapshot of intensity, field, and velocity for the 300 G cases for all investigated spectral types, illustrating the aforementioned points. We also show similar snapshots for the 100 G case and the 500 G case in Fig. \ref{fig:snap_100g} and Fig. \ref{fig:snap_500g}, respectively, for comparison. Plots for the other field strength cases are available in Appendix \ref{app:plots}. We note that despite imposing a moderate to strong vertical magnetic field in the various simulations, there are indications of weak, opposite polarity fields, especially at granular edges near the strong field concentrations (slight red areas in middle columns of Fig. \ref{fig:snap_300g} and subsequent snapshots). This is consistent with observations of solar plage regions \citep{buehler2015} and is a result of near-surface convection-driven recirculation of magnetic field, where the strong shear between up- and downflows churns and drags up field to the surface. On the other hand, there are also regions of mixed polarity further away from the strong flux concentrations, which could either be from recycled fields or from a SSD.

Despite the aforementioned effects on bolometric intensity in the presence of magnetic fields, the variation in the effective temperature calculated from the angle-averaged bolometric radiative flux is within $1\%$ of the reference SSD simulations (see Fig. \ref{fig:teff}). The trends in the changes of effective temperature with field strength are consistent with previous studies; for example, \citet[Table 1]{beeck3} and \citet[Table 2]{norris2023}. We note that the effective temperature calculated here assumes that the whole stellar surface has the same field strength. It is not meant to represent the actual effective temperature of a facular region on a star's surface, which would have a center-to-limb variation.

In addition to the previously established results in \citet{beeck3}, we report here the effect of faculae-like fields on thermodynamic structure and convection, and offer probable explanations for the changes observed compared to the reference SSD simulations. We take up the 300 G case simulations as a typical strong facular region for reference for these comparisons. Other field strengths show similar behavior. The magnitude of changes compared to SSD simulations depends on the field strength.

\subsection{Changes in the thermodynamic structure}\label{sec:res:td}

\begin{figure}[ht]
    \resizebox{\hsize}{!}
    {\includegraphics{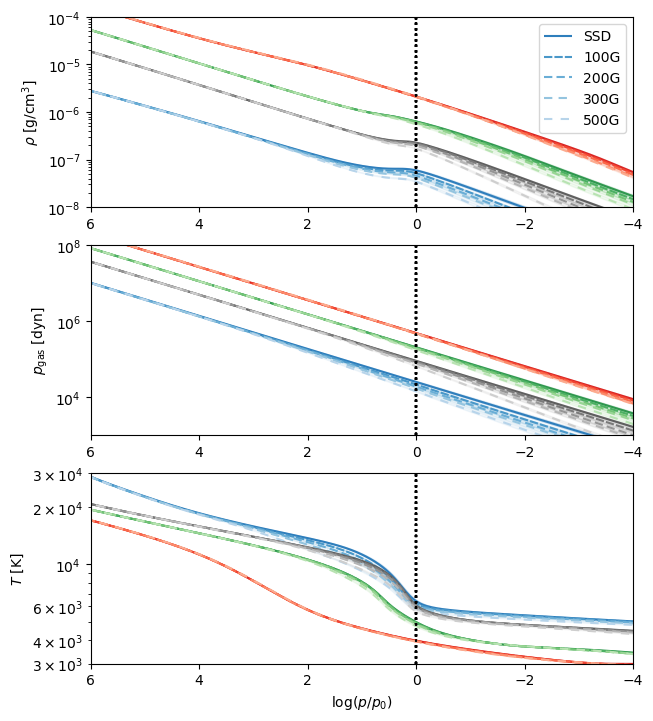}}
    \caption{Density, $\rho$ (\textit{top}), gas pressure, $p_{\rm gas}$ (\textit{middle}), and temperature, $T$ (\textit{bottom}), for all magnetic field strengths and stellar types. The horizontal axis is the number of pressure scale heights below the surface computed for the SSD case for each star (the left side of the plot is toward the bottom boundary and the right side is toward the top). The vertical dotted line marks the height at which $\avg{\tau}=1$ in each case.}
    \label{fig:td_all}
\end{figure}

\begin{figure}[ht]
    \resizebox{\hsize}{!}
    {\includegraphics{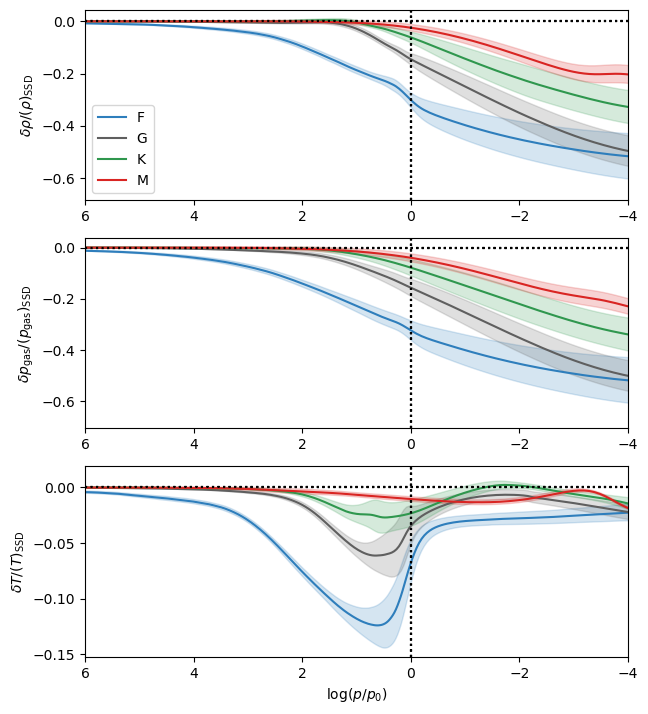}}
    \caption{Changes in density, $\rho$ (\textit{top}), gas pressure, $p_{\rm gas}$ (\textit{middle}), and temperature, $T$ (\textit{bottom}), for the 300 G case relative to the SSD case for all four stellar types in percent (\%). The error bands represent the $1\sigma$ variation in time for the horizontal means. All subsequent error bands have the same meaning.}
    \label{fig:td_dev}
\end{figure}

The horizontally averaged plots of density, temperature, and pressure show  changes relative to the SSD case, with the magnitude of change depending on the field strength (Fig. \ref{fig:td_all}). To examine the changes in detail, in Fig. \ref{fig:td_dev}, we plot the relative difference between the 300 G cases and their corresponding SSD runs for a given horizontally averaged quantity, $q$, as $(q-q_{SSD})/q_{SSD}$ against a pressure scale axis, $\log(p/p_0)$, as described in Sec. \ref{sec:methods}. We computed the deviations on a geometric height axis and used the SSD $\log(p/p_0)$ axis to plot them. We see that both the density, $\rho$, and gas pressure, $p_{\rm gas}$, are reduced around the stellar surface in the facular simulation for all the stars, with the magnitude and depth of change somewhat proportional to $T_{\rm eff}$. Near the bottom of the box (left side of the plots), the changes are negligible. The temperature profile shows a similar dip below the surface ($\log(p/p_0) \sim 0.5-1$), but at the surface (dotted vertical lines), the change becomes almost negligible. Possible explanations for this trend are discussed in Sec. \ref{sec:disc}.

\begin{figure}[ht]
    \resizebox{\hsize}{!}
    {\includegraphics{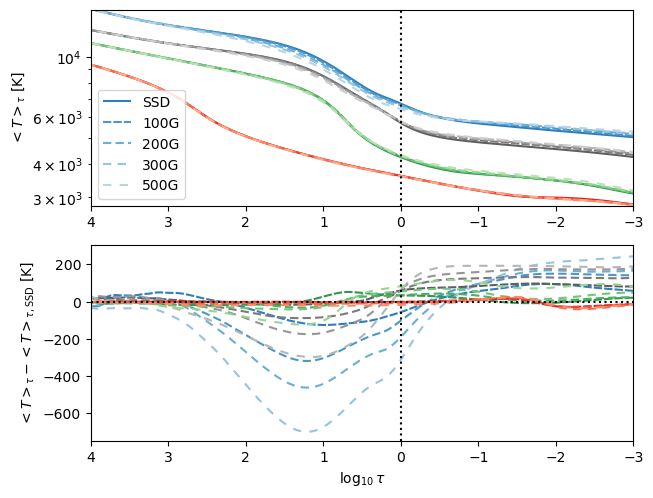}}
    \caption{Temperature, $T$, averaged over equal Rosseland tau $\tau$ surfaces instead of geometric horizontal slices for each stellar and magnetic field case. Absolute profiles (\textit{top}) and difference in the imposed field temperature profiles relative to the SSD cases (\textit{bottom}).}
    \label{fig:t_tau}
\end{figure}

We note that the bottom panel of Fig. \ref{fig:td_dev} seems to indicate that the average temperature in faculae is always lower than the corresponding SSD temperature, even above the optical surface ($\log(p/p_0)<0$), which seems to contradict solar observations \citep[Table 2]{Solanki1986,Solanki1992,buehler2019}. This is a consequence of horizontal averaging over geometric slices. The observed temperatures are averages over $\tau$ surfaces. In the top panel of Fig. \ref{fig:t_tau}, we take the full 3D temperature cubes from single snapshots for each star and field case, and plot the average of surfaces of equal Rosseland optical depth, $\tau$. The change in the temperature profiles (Fig. \ref{fig:t_tau}, bottom panel) shows that the F, G, and K stars have a higher average temperature above $\log_{10}\tau \sim -1$, with the magnitude of change somewhat proportional to field strength and $T_{\rm eff}$. We also note that the enhancement in temperature relative to the SSD case seems to begin deeper in the atmosphere for cooler stars. The M-star simulations do not show any such trend and the difference in the $\tau$ stratification of temperature between the different simulations is rather small. For the hotter stars, the trend is consistent with solar observations.

\subsection{Changes in the velocities}\label{sec:res:vels}

\begin{figure}[ht]
    \resizebox{\hsize}{!}
    {\includegraphics{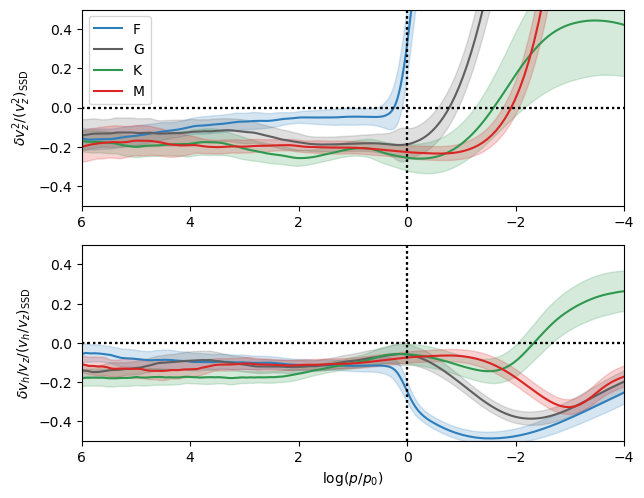}}
    \caption{Changes in square of vertical velocity, $v_z^2$ (\textit{top}), and in the ratio of horizontal to vertical velocities, $v_{h,rms}/v_{z,rms}$ (\textit{bottom}), for the 300 G case relative to the SSD case for all four stellar types.}
    \label{fig:vel}
\end{figure}

The presence of a magnetic field is known to suppress convective velocities in general \citep{WeissProctor2002}. This is usually attributed to the Lorentz force acting on plasma motions through the $(v\times B)\times B$ term in the MHD momentum equation and quenching the overturning motion. In Fig. \ref{fig:vel}, we see the expected decrease in convective velocities (considered here in terms of $v_z^2$) for all the stars, with the decrease roughly around 20\%, all the way down to the bottom of the box. Near the surface, the trends are reversed. Here, since the mode of energy transport changes from convective to radiative, the interpretation of the increase in $v_z^2$ above the surface is not so straightforward. The ratio of horizontal to vertical velocities shows a small decrease in the deeper convection, indicating a corresponding decrease in horizontal velocities as well. This is well correlated with a decrease in the horizontal extent of convection in the presence of magnetic fields \citep[ Sec. 4.1]{paper3}.

\subsection{Bolometric intensity}\label{sec:res:intensity}

\begin{figure}[ht]
    \resizebox{\hsize}{!}
    {\includegraphics{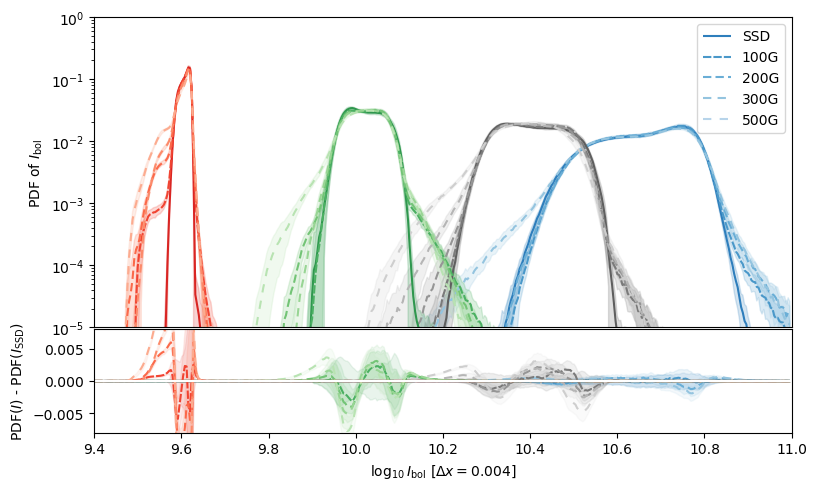}}
    \caption{PDF of logarithmic bolometric intensity for all cases (\textit{top}) and the difference in PDFs between the imposed field cases and the SSD case (\textit{bottom}).}
    \label{fig:ipdf}
\end{figure}

The PDF of disk-center bolometric intensity in Fig. \ref{fig:ipdf} shows trends consistent with previous simulations of near-surface stellar convection with faculae-like fields \citep{salhab2018,beeck4}. Briefly, the distributions can be interpreted in terms of two peaks, corresponding to granules and intergranular lanes. The presence of magnetic fields adds to the dark (left) flank through formation of dark pore-like structures. This effect increases with increasing field strength and is strongest for the M star. For the K-star (and, to a lesser degree, the other stars), the bright (right) flank is also affected by the formation of bright points and bright filigree in the intergranular lanes. More generally, the shape of the two peaks also seems to be slightly affected, with the average granular brightness slightly decreasing (peak shifting to the left, visible in the residuals plot below as an increase in density on the left side of the peak granule intensity and a decrease on the right side). We note that these results hold only for the bolometric intensity at the disk center and the effect of magnetic fields on intensity distribution at different viewing angles as well as in different spectral ranges is expected to be different \citep{norris2017,norris2023}. We plan to investigate this more generally in a subsequent paper.

\subsection{Morphology of magnetic field concentrations and associated flows at and below the stellar surface}\label{sec:res:bzvz}

The simulations exhibit a diversity of flux tube and flux sheet sizes as well as magnetic field strengths. To examine these further, we plot the horizontal and vertical cuts of $B_z$ and $v_z$ in Fig. \ref{fig:bzvz_300g} for all the simulations with 300 G imposed magnetic field.  We note that, for the F star, most of the magnetic field seems to concentrate into a wide and vertically contiguous intergranular lane going all the way to the bottom boundary (for the $z=8.53$ Mm slice, at $x=10$ Mm and $y=13.5$ Mm, roughly), whereas for the cooler stars, the lanes become narrower and more spread out. For all the stars, most of the magnetic field tends to collect in the intergranular lanes that are contiguous almost all the way to the bottom of the simulation domain. The red contours in the field maps in Fig. \ref{fig:bzvz_300g} enclose regions where $B_z > B_{\rm eqp}$, with $B_{\rm eqp} = \sqrt{8\pi \avg{p_{\rm gas}}}$ (with averaging done over the horizontal simulation domain). The blue contours enclose regions where the magnetic energy in $B_z$ is stronger than the average kinetic energy at a given height. We note that the magnetic field for the F star stays relatively strong compared to the gas pressure for the upper part of the box, whereas for the M star this is true only near the surface. For all stars, however, the magnetic energy remains stronger than the kinetic energy in magnetic concentrations throughout the simulated convection zone. We also note that the magnetic field concentrations become quite fragmented in the lower half of the simulation boxes for all stars.

\begin{figure*}[ht]
    \centering
    \resizebox{\hsize}{!}
    {\includegraphics[width=17cm]{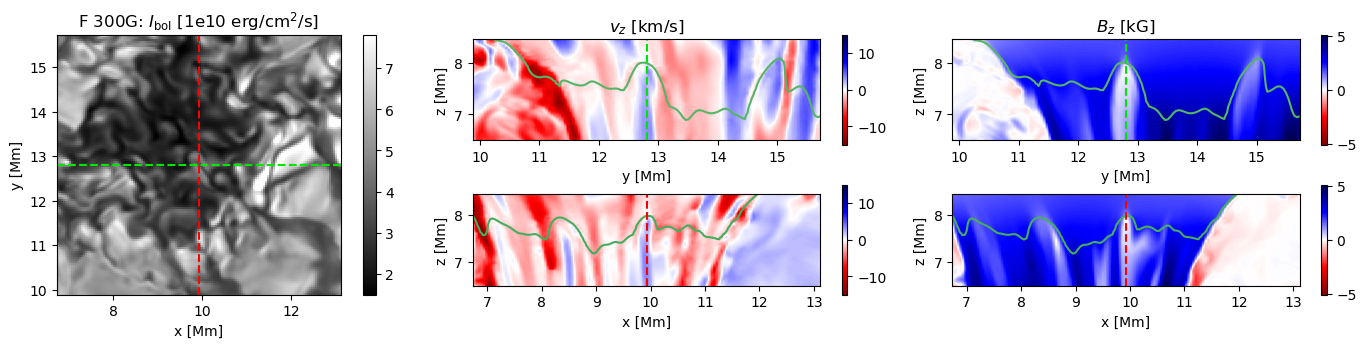}}
    \caption{Zoom-in of the F-star pore-like structure in Fig. \ref{fig:snap_300g} showing umbral-dot-like features. Bolometric intensity of the zoomed-in region (\textit{left}) with vertical cuts in convective velocity, $v_z$ (\textit{middle}), and the vertical magnetic field, $B_z$ (\textit{right}), along the $y$ and $x$ axes. The dashed colored lines represent the locations of the cuts and the solid green line represents the $\tau=1$ surface. The bottom of the simulation box corresponds to $z=0$ for reference.}
    \label{fig:ud}
\end{figure*}

In the velocity cuts, downflows are associated with the edges of field concentrations. We note that, even in the large field concentrations, the $v_z$ flow maps show a network of small-scale motions that seem convective in nature. This is especially apparent for the F star. The corresponding bolometric intensity map for the F star (Fig. \ref{fig:snap_300g}, top left panel) also shows intensity variations somewhat correlated with these velocity structures. These flows are a version of flows associated with umbral dots and are, in general, a signature of convection in strong vertical fields \citep{narayan2010}. A zoom in of the F-star pore-like feature is presented in Fig. \ref{fig:ud}, showing multiple clear cusp-like structures in upflows embedded in a network of relatively weak downflows. The magnetic field structure shows a weakening of field in the cusp, compared to the surroundings. These features are correlated with upward bumps in the optical surface and bright features in the bolometric intensity, as was seen in previous simulations \citep{muram_umbra_2006}, as well as observations \citep{tino2008} of umbral dots.

\subsection{PDFs of surface magnetic fields}\label{sec:res:bpdf}

\begin{figure}[ht]
    \resizebox{\hsize}{!}
    {\includegraphics{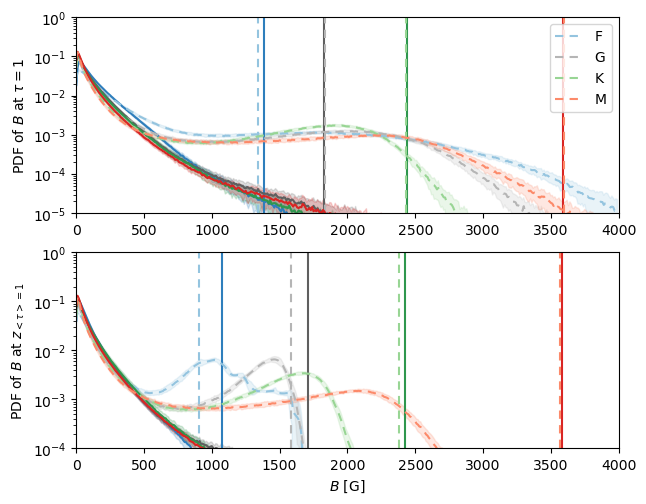}}
    \caption{\textit{Top:} PDFs of the strength of magnetic field $B$ for the 300 G (\textit{dashed}) and the SSD (\textit{solid}) cases, calculated for the $\tau=1$ surface. \textit{Bottom:} Same as above, but calculated for the geometric surface $z_{\avg{\tau}=1}$ corresponding to the height at which $\avg{\tau}=1$. The vertical lines correspond to the pressure equipartition field, $B_{\rm eqp}$, for both field cases for all stars, calculated as described in \citet[Fig. 2 and Sec. 4.1]{paper3}.}
    \label{fig:bpdf}
\end{figure}

The PDFs of the magnitude of magnetic field at the optical ($\tau = 1$) surface (Fig. \ref{fig:bpdf}, top panel) show a rather similar distribution for the facular simulations in the hecto-gauss (hG) range, with a slight trend with $T_{\rm eff}$ (hotter stars have slightly higher fraction of stronger hG fields). For comparison, the PDF of SSD simulations are plotted as solid lines. They show the same trend until roughly 500 G. For higher values they diverge from the facular fields. In the kilogauss (kG) range, the distribution of the 300 G simulations is similar till roughly 2.5 kG. The strongest fields seem to follow a trend (hotter stars have higher fraction of stronger fields) with stellar type, except for the M star. All stellar types with an imposed field seem to show field strengths in excess of the corresponding gas plus hydrodynamic turbulent pressure equipartition value \citep{paper3}. We note that inferring anything about the relation between field strength and pressure on a $\tau$ surface is difficult since the surface itself is corrugated, with stronger fields lying geometrically deeper in intergranular lanes due to evacuation of plasma and associated opacity changes.

To compare against gas pressure, we must look at the PDF of fields at the geometric surface corresponding to $\avg{\tau}=1$ (bottom panel), the idea being that at any given height, the simulation box is expected to be in horizontal pressure balance on average. Here we see that the kG fields show a distinct peak for the imposed field cases, with the peak at the pressure equipartition field strength for the F and G star. For the K and M star, the peaks are less pronounced and lie at decidedly sub-equipartition values. The peak field strength is related to the efficiency of the convective collapse mechanism \citep{spruit_linear_1979}. Cooler dwarfs with higher surface gravity tend to be more stable against convective collapse and are, accordingly, less likely to develop field concentrations with equipartition field strength \citep{rajaguru2002_stellar_cc}. Coming back to the top plot, we note that stronger fields are observed in this plot because they are observed at a deeper geometrical depth, where $p_{\rm gas}$ is higher than at $z_{\avg{\tau}=1}$.

\section{Discussion}\label{sec:disc}

\subsection{Factors affecting the temperature stratification}\label{sec:disc:tem}

\begin{figure}[ht]
    \resizebox{\hsize}{!}
    {\includegraphics{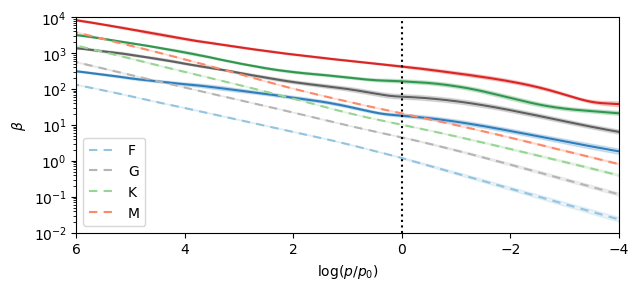}}
    \caption{Plasma $\beta$ as a function of $\log(p/p_0)$ for the 300 G (\textit{dashed}) and SSD (\textit{solid}) cases for all stars.}
    \label{fig:beta}
\end{figure}

The presence of magnetic fields primarily affects the thermodynamics by introducing an additional component to the horizontal pressure balance from magnetic pressure $p_{\rm mag} = B^2/8\pi$. The main consequence of the first effect is a reduction in gas pressure at the surface. In the regions where $\beta = p_{\rm gas}/p_{\rm mag} \sim 1-10$, this effect becomes significant. In Fig. \ref{fig:beta}, we see that for the 300 G case for the F star, this is the case near the optical surface. The decrease in $p_{\rm gas}$ is simply balanced by extra $p_{\rm mag}$ to maintain horizontal pressure balance. For the cooler stars, since $\beta$ is significantly higher at the same depths, this effect is quite weak. The changes in density largely follow the changes in pressure.

\begin{figure}[ht]
    \resizebox{\hsize}{!}
    {\includegraphics{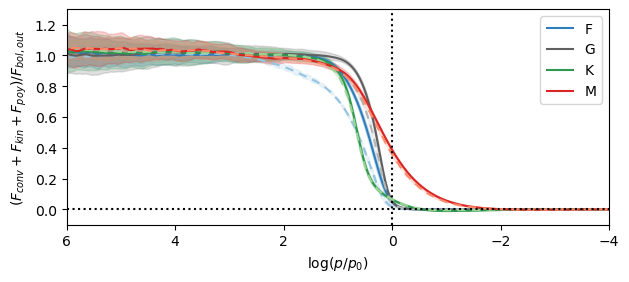}}
    \caption{Fractional non-radiative flux, $F$; that is, sum of the convective $F_{\rm conv}$, kinetic $F_{\rm kin}$, and Poynting $F_{\rm poy}$ flux normalized by the angle-averaged outgoing bolometric flux for the 300 G (\textit{dashed}) and SSD (\textit{solid}) cases.}
    \label{fig:flux}
\end{figure}

In addition to the pressure changes, magnetic field also restricts convective velocities, as is noted in Sec. \ref{sec:res:vels}. Consequently, this is expected to reduce the convective flux reaching the surface. The picture is not so simple, however. The total flux has four components: convective ($F_{\rm conv}$), kinetic ($F_{\rm kin}$), Poynting ($F_{\rm poy}$), and radiative ($F_{\rm rad}$) flux (see Appendix \ref{app:fluxes} for details). Below the optical surface, $F_{\rm conv}$ is always positive, $F_{\rm kin}$ and $F_{\rm poy}$ are negative on average, and $F_{\rm rad}$ is negligibly small \citep{beeck1}. Near the surface, the mode of energy transport changes from convective to radiative, and the sum of first three terms goes to zero. As the magnetic field evacuates plasma, the drop in density also causes a drop in opacity. This allows radiation to escape from deeper layers. In Fig. \ref{fig:flux}, it can be seen that this effect is the strongest for the hottest star in the 300 G case, with a noticeable decrease in the total non-radiative flux (relative to the SSD case) already visible up to 2 pressure scale heights below the surface. As the deeper regions lose energy, they become cooler compared to the SSD case (by up to 15\% in the F-300 G case). At the surface, however, this difference becomes small again.

\begin{figure}[ht]
    \resizebox{\hsize}{!}
    {\includegraphics{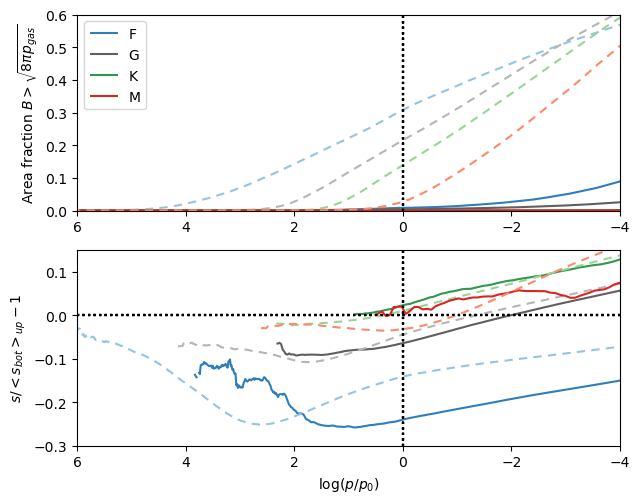}}
    \caption{Area fraction of regions with $B>\sqrt{8\pi p_{\rm gas}}$ (\textit{top}) and the corresponding entropy normalized by the upflow entropy at the bottom boundary, $s/\avg{s_{\rm bot}}_{\rm up}$, in the strong-field regions calculated from single cubes for the 300 G (\textit{dashed}) and SSD (\textit{solid}) cases.}
    \label{fig:strongB_entropy}
\end{figure}

\begin{figure}[ht]
    \resizebox{\hsize}{!}
    {\includegraphics{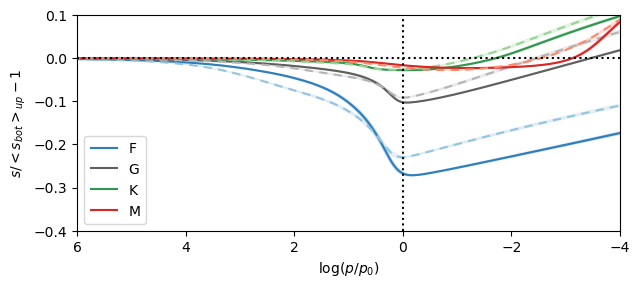}}
    \caption{Horizontally averaged entropy normalized by the upflow entropy at the bottom boundary, $s/\avg{s_{\rm bot}}_{\rm up}$, for the 300 G (\textit{dashed}) and SSD (\textit{solid}) cases.}
    \label{fig:avg_entropy}
\end{figure}

The fact that the mode of energy transport starts transitioning from convective to radiative at a larger depth for the 300 G case implies that the plasma in strong field regions becomes stably stratified, i.e., sub-adiabatic, with a positive entropy gradient. In Fig. \ref{fig:strongB_entropy}, we plot the area fraction of regions where magnetic pressure is stronger than the local gas pressure and the corresponding strong-field average entropy in these regions. We contrast these profiles to the horizontally averaged entropy profiles for the entire time series in Fig. \ref{fig:avg_entropy}. It is quite apparent that the regions with strong fields occupy a significant area fraction near the surface for the F-300 G case and the corresponding entropy profile has the entropy minimum located deeper in the convection zone compared to the average profile. For the F-SSD case, the area fraction is quite small so the contribution from these regions is negligible. Other stars show a similar trend, but to a lesser degree. This effect is the same as that seen in average entropy profiles of the umbral region in starspot simulations \citep[Fig. 5]{starspots2025}

The complete picture then is as follows: the entropy minimum of the strong-field regions in the facular simulation is located geometrically deeper compared to the weak-field regions. This results in a decrease in the horizontally averaged temperature below the surface. Above the optical surface of these strong-field regions, the sub-adiabatic (stable) stratification causes the temperature profile to be flatter than expected when compared to the surrounding weak-field convecting medium. In addition, there is a probable contribution from the hot walls of the surrounding convecting plasma. When compared to the respective SSD simulation, these effects together result in a temperature profile for the facular simulation that is cooler below the optical surface (by up to 15\% for the F-300 G simulation) but rather similar above the surface (a change of <5\%).

\subsection{Factors affecting the surface magnetic field distribution}\label{sec:disc:field}

\begin{figure}[ht]
    \resizebox{\hsize}{!}
    {\includegraphics{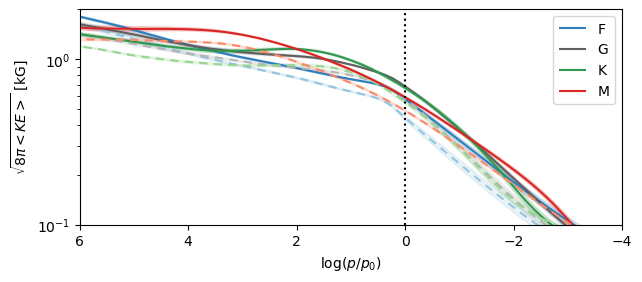}}
    \caption{Kinetic energy equipartition field strength as a function of pressure scale height for the 300 G (\textit{dashed}) and SSD (\textit{solid}) cases.}
    \label{fig:ke_eqp}
\end{figure}

The magnetic field is, in general, concentrated in downflows, as flux expulsion from convective flows \citep{parker1963,weiss1996_fe} drives the magnetic field into the intergranular lanes. At the fixed geometrical depth corresponding to the average optical surface, the characteristic field strength of these concentrations is lower for hotter stars (peaks in Fig.~\ref{fig:bpdf}, bottom panel). This is a consequence of horizontal balance against ambient gas pressure, which decreases with $T_{\rm eff}$. Since the total magnetic flux must remain conserved, this is consistent with strong fields at the same depth occupying a larger area fraction for hotter stars (see Fig.~\ref{fig:bzvz_300g}). Deeper in the simulation boxes, the field strength falls significantly below the pressure equipartition value and the spatial distribution no longer follows the trend with effective temperature. The blue outlines delineating contours of fields comparable to kinetic equipartition look rather similar between the different stars. This is due to the fact that the mean kinetic energy equipartition value of magnetic fields is rather similar with depth, independent of stellar type or imposed field (Fig. \ref{fig:ke_eqp}).

We note that the field strengths achievable are limited by equipartition with the plasma pressure, which is lower for hotter stars. For the F and (to some extent) the G stars, pressure equipartition values are actually achieved in the strongest flux concentrations (red contours). In the cooler K and M stars, however, the field at the surface is sub-equipartition. This is consistent with the result of \citet{rajaguru2002_stellar_cc}, who find that the minimum critical plasma beta ($\beta=\beta_c$) for which convective collapse can be triggered is higher for cooler stars. The typical maximum $\beta$ can be estimated from the constraint that a flux tube will only form if the initial field is weak compared to plasma flows. Following the same kinetic energy equipartition arguments as in Sec. 4.2 of \citet{spruit_saturated_1979}, the maximum $\beta$ in a flux tube must be less than $2/(\gamma M^2)$, where $\gamma$ is the adiabatic index and $M$ is the Mach number. For a rough estimate, we used the values from Fig. 6 in \citet{beeck1} for $M$ ($\sim 0.22$ for the M star and $\sim 1$ for the F star) and took $\gamma$ as 1.3 for the F star and 1.66 for the M star \citep[see Table 1]{rajaguru2002_stellar_cc}. With these values, maximum $\beta$ varies from $\sim 1.5$ for the F star to $\sim 25$ for the M star. The critical $\beta_c$ values were estimated from a bi-variate spline fit of Table 2 of \citet{rajaguru2002_stellar_cc}, giving $\beta_c \sim 0.026$ for the F star and $\beta_c \sim 8.756$ for the M star.

To estimate the saturated state of a flux tube undergoing convective collapse on different stars, we assumed that the relation between the initial and the final $\beta$ for such a flux tube is relatively constant and has the same slope as in Fig. 2 of \citet{spruit_saturated_1979}; that is, $\beta_{\rm sat.} \sim \beta_{\rm initial}^a$, with $a\sim -1/3$. Since $\beta_c$ is different for different stars, the relation can be written as
\begin{align}
    \log\beta_{\rm sat.} & \approx (1-a)\log \beta_c +a\log \beta_{\rm initial}
.\end{align}

In the extreme case in which the flux tube is completely evacuated ($\beta_{\rm sat.} \to 0$), the maximum field strength is roughly the gas pressure equipartition value. Hence, the observed field strength can be estimated as $B_{\rm eqp}/\sqrt{1+\beta_{\rm sat.}}$. The minimum achieved value of $\beta_{\rm sat.}$ determines the observed field strength. Combining the estimated $\beta_{\rm sat.}$ values computed using the above expression and the $B_{\rm eqp}$ values from Fig. \ref{fig:bpdf} bottom panel, we obtain the maximum field strength of $\sim 1$ kG for the F star and $\sim 1.3$ kG for the M star. Despite the surface gas pressure of the M star being almost an order of magnitude higher than that of the F star, the field strengths achieved are quite similar and in line with Fig. \ref{fig:bpdf}. This rough analysis highlights the importance of considering the efficiency of the convective collapse mechanism for estimating small-scale magnetic field strength on stellar surfaces. We note that we have neglected the effect of the depth of the Wilson depression on the observed field strength, since it is a nontrivial function of opacity, density, and temperature stratification in the flux tube.

The trends in disk-center bolometric intensity and surface field strength distribution (see Sec. \ref{sec:res:intensity}) are largely similar to those obtained in previous such studies, such as those of \citet{beeck3} and \citet{salhab2018}. We note the remarkable similarity of field strength distribution at the optical surface for all the stellar types considered. The hG field strength distribution is expected to depend on the kinetic energy distribution because of a rough balance between kinetic and magnetic energy in the plasma flow. Since the surface kinetic energy is roughly similar at all scales for the stellar types considered \citep[see Fig. 7]{paper3}, it is to be expected that the magnetic field distribution is also rather similar in this range.

The observed kG field values at the surface also depend on their Wilson depression. The depth of the Wilson depression depends on the opacity and density of the plasma in the intergranular lanes (where the opacity itself depends largely on temperature). If we assume that a magnetic flux tube is in temperature equilibrium with its surroundings, we can take $\beta$ to be constant. In this case, a deeper Wilson depression corresponds to a a higher observed field strength. For an F star, since the saturated field strength is basically at pressure equipartition, the degree of evacuation is quite high, and the observed field is in equipartition with gas pressure at a larger depth. This is not the case for the M star, where the degree of evacuation is sub-equipartition, even for the strongest possible field from the discussion in the previous paragraph. The field strengths actually observed end up being quite similar between the stellar cases considered in this paper. We note that this is a qualitative explanation. A possible analysis in the framework of thin flux tube and/or sheet approximation was done previously by \citet{YellesChaouche2009}.

\section{Conclusion} \label{sec:conc}

In this study, we investigate the effects of facular magnetic fields on thermodynamic structure, the emergent bolometric intensity, and the distribution of field strengths at the optical surface of cool main-sequence stars ranging from F to M spectral type. We find that, as was expected, strong magnetic fields reduce the density and pressure near the surface due to plasma evacuation in the presence of magnetic fields \citep[see Chapter 8]{stix1989}. We also see a change in the entropy (temperature) structure, indicating a transition to radiative energy transport in strong-field regions deeper in the convection zone as compared to surrounding weak-field regions. While there is a decrease in average temperature just below the surface, the temperature profile above the surface is rather similar to the corresponding SSD simulation. 

The morphology of magnetic features throughout the simulated convection zone show a larger and more contiguous area fraction of strong fields for the hotter stars for a variety of imposed average field strengths (Fig. \ref{fig:snap_300g},\ref{fig:snap_100g},\ref{fig:snap_500g}). This is a consequence of lower ambient pressure on hotter stars. The distributions of field strengths show peaks for strong fields at different values for different stellar types on a constant geometric slice corresponding to the optical surface. When the corrugation of the optical surface is accounted for, the distributions appear rather similar for strong field (> 1 kG) values, with the F star showing a slightly higher fraction of strong fields. The largest concentrations on the F and G stars show similarities to solar pores. The convective velocity and the magnetic field structure in the pore-like regions in the F star shows similarities to the convective velocities and the magnetic field structure of solar umbra dot simulations.

The distributions of intensities show clear extended bright and dark flanks in the $I_{\rm bol}$ PDF for almost all simulations with magnetic fields, compared to the reference SSD case (Fig. \ref{fig:ipdf}). The bright flank remains rather similar for different field strengths for a given stellar type, whereas the dark flank becomes stronger with increasing field strength. The simulations exhibit a preference for darker features with decreasing $T_{\rm eff}$.

Magnetic fields affect not just the emergent intensity but also the structure and convective properties of plasma in the convection zone. The rapidly expanding field of exoplanet characterization requires precise and accurate modeling of the stellar signal, which itself depends on detailed and accurate simulations of near-surface stellar magnetism. This study aims to fill that gap.

\begin{acknowledgements}
    The authors would also like to thank Yvonne Unruh for insightful discussions and general help in improving this manuscript. The authors would like to thank the anonymous referee for their careful review, which helped improve the presentation of results in this manuscript. TB is also grateful for access to the supercomputer Cobra at Max Planck Computing and Data Facility (MPCDF), on which all the simulations were carried out. This project has received funding from the European Research Council (ERC) under the European Union’s Horizon 2020 research and innovation program (grant agreement No. 695075 and No. 101097844). VW and AIS acknowledge support from the European Research Council (ERC) under the European Union’s Horizon 2020 research and innovation program (grant no. 101118581). TB and AIS also acknowledge support from DFG grant SH1489/1.
\end{acknowledgements}

\bibliography{biblio}{}
\bibliographystyle{aa}

\begin{appendix}

\section{Fluxes}\label{app:fluxes}
The various fluxes mentioned in Sec. \ref{sec:disc} are expressed as,
\begin{align}
    F_{\rm conv} &= v_z(E+p) \label{eqn:fconv} \\
    F_{\rm kin} &= v_z(\rho v^2/2) \label{eqn:fkin} \\
    F_{\rm poy} &= v_z (B_x^2 + B_y^2) - B_z (B_x v_x + B_y v_y) \label{eqn:fpoy}
\end{align}
Here, $E$ is the internal energy, $p$ is the gas pressure, and the other symbols have their usual meanings ($v$ for velocity and $B$ for magnetic field). In the ideal scenario, the horizontally averaged vertical momentum flux in the simulation domain $\rho v_z$ should be zero. However, due to the presence of box modes, this term is, in general not zero at all time steps. This affects the calculation of the convective (Eq. \ref{eqn:fconv}) and kinetic (Eq. \ref{eqn:fkin}) fluxes. Even though the average over time averages out the box mode, the calculation of standard deviation over time is misleading in its magnitude. To account for the contribution of the box mode, the net convective flux is calculated as,
\begin{align}
    \avg{F_{\rm conv}} &= \avg{v_z(E+p)} - \avg{\rho v_z}\avg{(E+p)/\rho}\\
    \avg{F_{\rm kin}} &= \avg{v_z(\rho v^2/2)} - \avg{\rho v_z}\avg{v^2/2}
\end{align}
The Poynting flux is not strongly affected by this term since there is no $\rho$ component in its expression.

\section{Opacity binning}\label{app:opac}

The procedure leading to the selected opacity bins, briefly, is as follows: the formation height (that is, the value of $\tau_{\rm ref}$ where $\tau_{\lambda}=1$ for each $\lambda_i$) for each wavelength point in the opacity distribution function (including continuum opacity) is calculated from a 1D atmosphere, which itself is computed from averaging a simulation timeseries horizontally and temporally. These opacity points are then averaged (either as Rosseland mean or Planck mean, depending on whether they form lower or higher in the atmosphere, respectively) over each bin to create a multibin opacity table. The simulation is run further with the new opacity table for enough time for most of the atmosphere to adjust to the new opacities (a couple of granulation timescales).

The distribution of wavelength points between different opacity bins depends on the mean stratification. Some points may fall in a different bin if the mean stratification changes, which can change the average opacity of that bin. The mean stratification couples back to the radiation via the radiative heating/cooling term in the energy equation. To make sure that the stratification that is used to compute the opacity bins is consistent with the structure on which the radiative scheme actually operates on, the procedure is iterated by using the evolved atmosphere (averaged horizontally and over 1-2 granulation timescales) at the end of each previous iteration. The changes in the effective temperature and thermodynamic structure near the surface remain small between the iterations and rapidly converged within 2-3 iterations per star.

\section{Additional figures} \label{app:plots}

\begin{figure*}[ht]
    \centering
    \includegraphics[width=17cm]{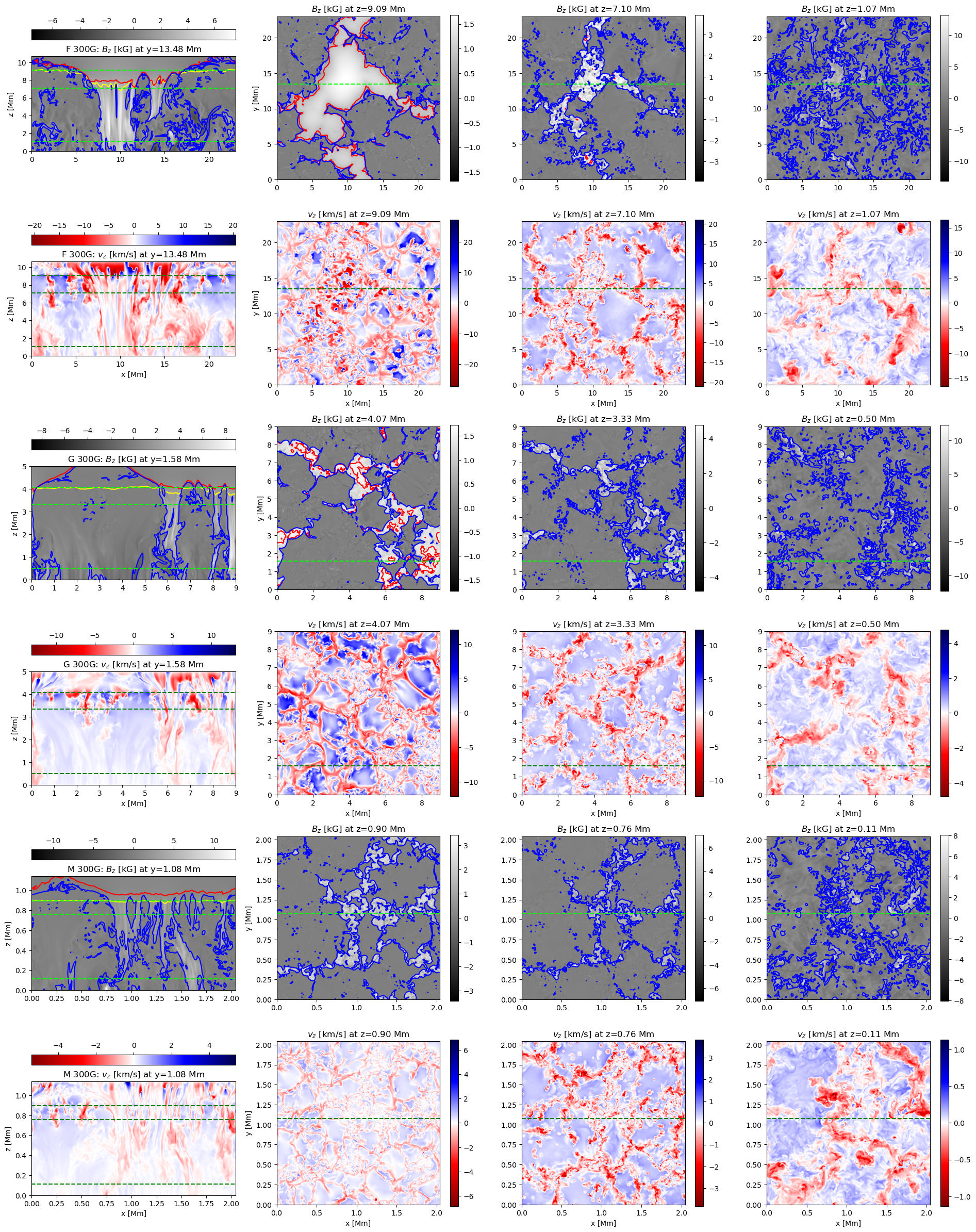}
    \caption{Geometric cuts of $B_z$ (\textit{odd rows}) and $v_z$ (\textit{even rows}) for the 300 G simulations. First column shows vertical cuts and the columns to the right show horizontal cuts. The locations of the cuts are specified in dashed green lines. The yellow contour is the surface corresponding to $\tau=1$. The horizontal cuts in the second column correspond to $z_{\avg{\tau}=1}$. The red contours enclose regions with $B_z>\sqrt{8\pi \avg{p_{\rm gas}}}$. The blue contours enclose regions where $B_z>\avg{\rho v^2/2}$. Here, the averages $\avg{p_{\rm gas}}$ and $\avg{\rho v^2/2}$ are computed over the horizontal slice being plotted. The rows correspond, from top to bottom, to F-, G- and M-star, respectively.}
    \label{fig:bzvz_300g}
\end{figure*}

\begin{figure*}[ht]
    \centering
    \includegraphics[width=17cm]{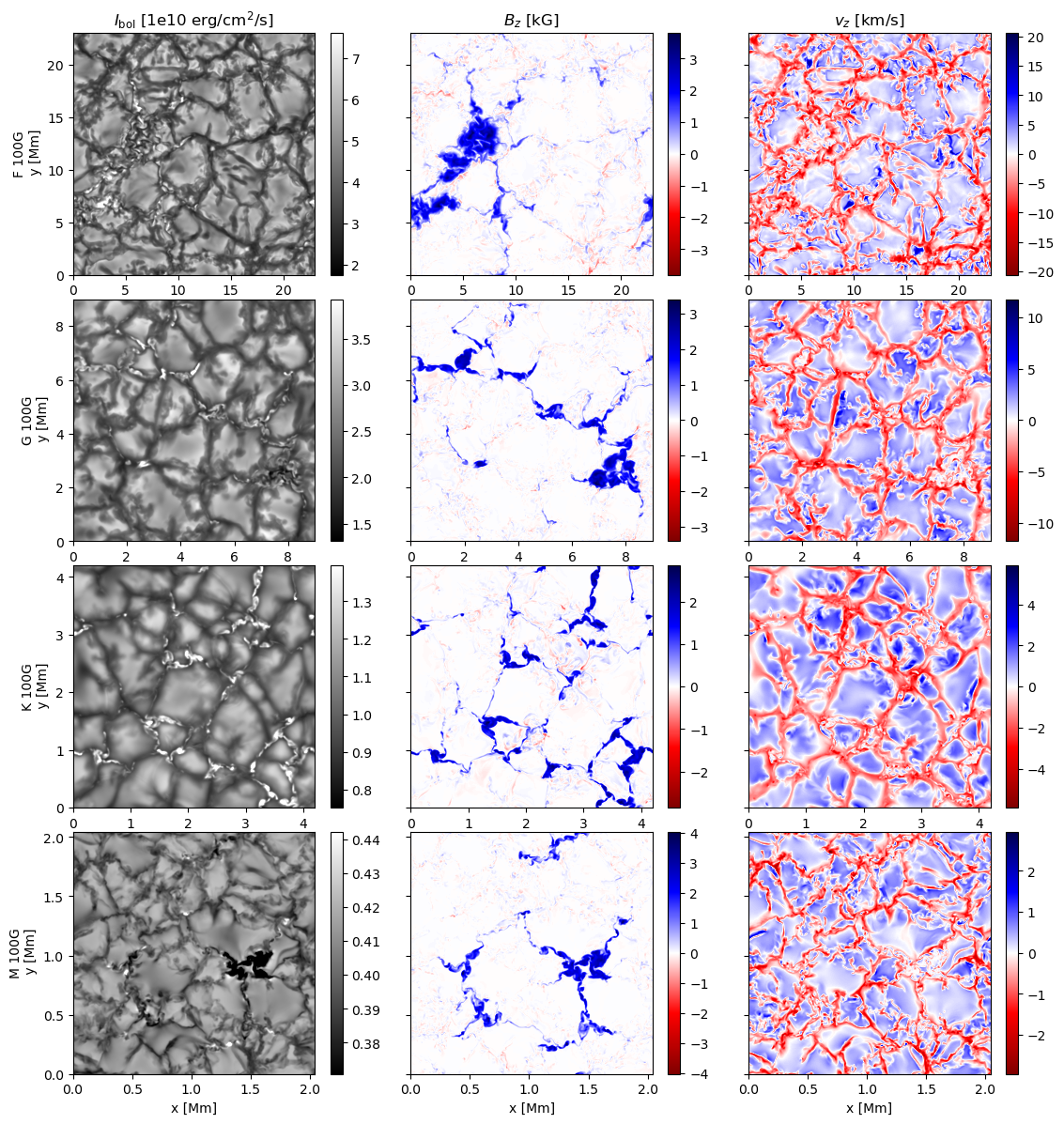}
    \caption{Same as Fig. \ref{fig:snap_300g}, but for the 100 G cases.}
    \label{fig:snap_100g}
\end{figure*}

\begin{figure*}[ht]
    \centering
    \includegraphics[width=17cm]{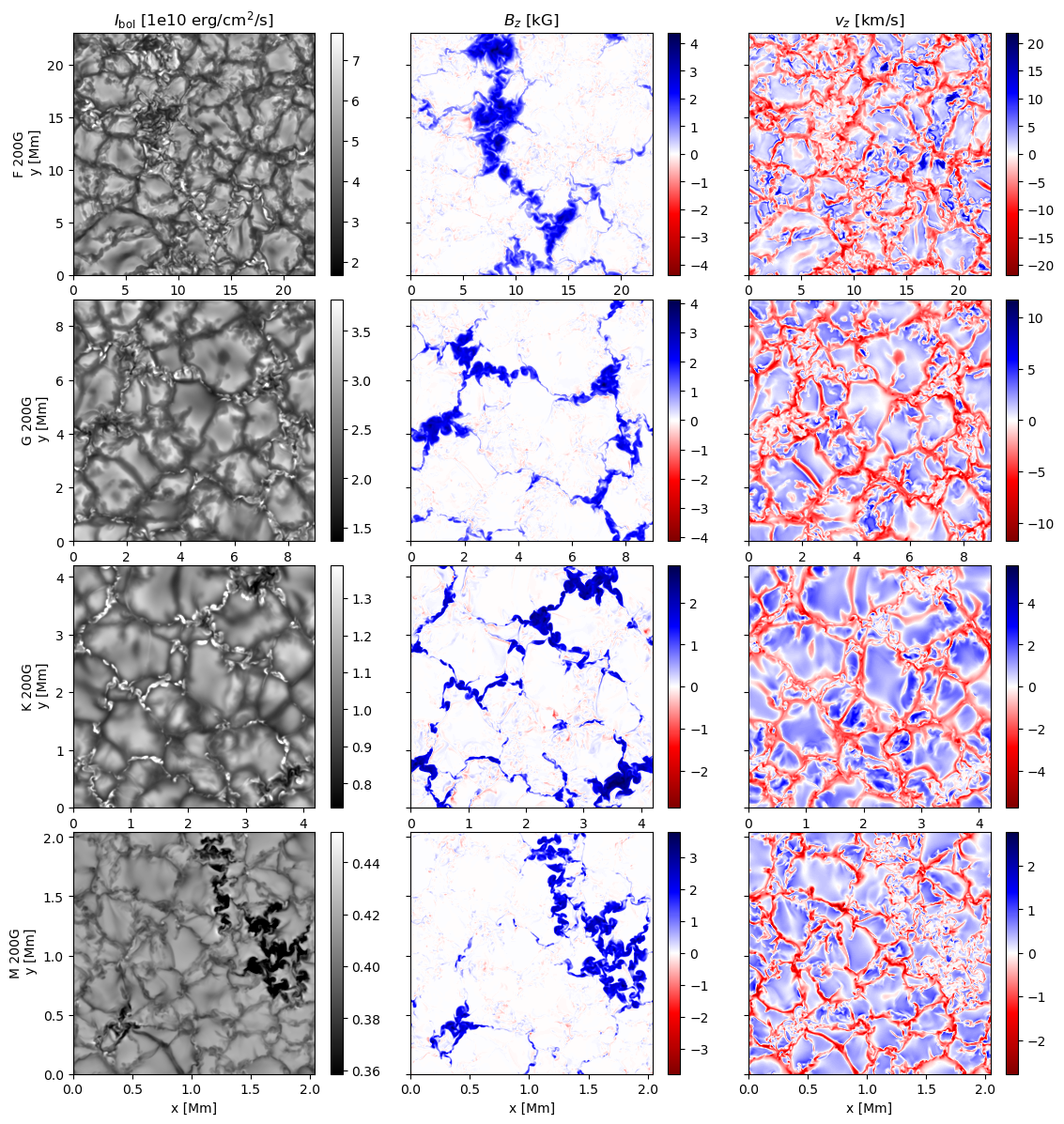}
    \caption{Same as Fig. \ref{fig:snap_300g}, but for the 200 G cases.}
    \label{fig:snap_200g}
\end{figure*}

\begin{figure*}[ht]
    \centering
    \includegraphics[width=17cm]{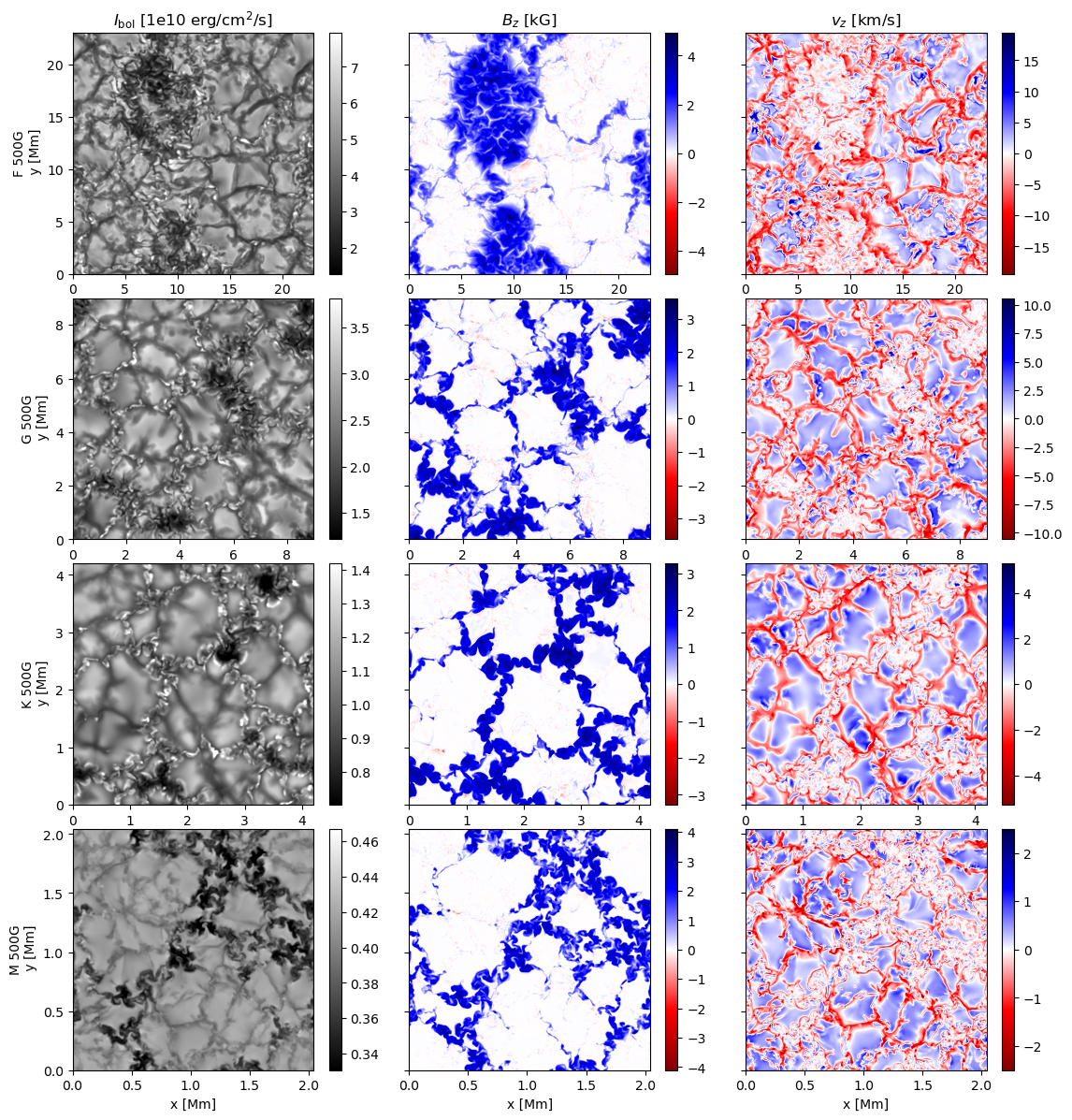}
    \caption{Same as Fig. \ref{fig:snap_300g}, but for the 500 G cases.}
    \label{fig:snap_500g}
\end{figure*}

\begin{figure*}[ht]
    \centering
    \includegraphics[width=17cm]{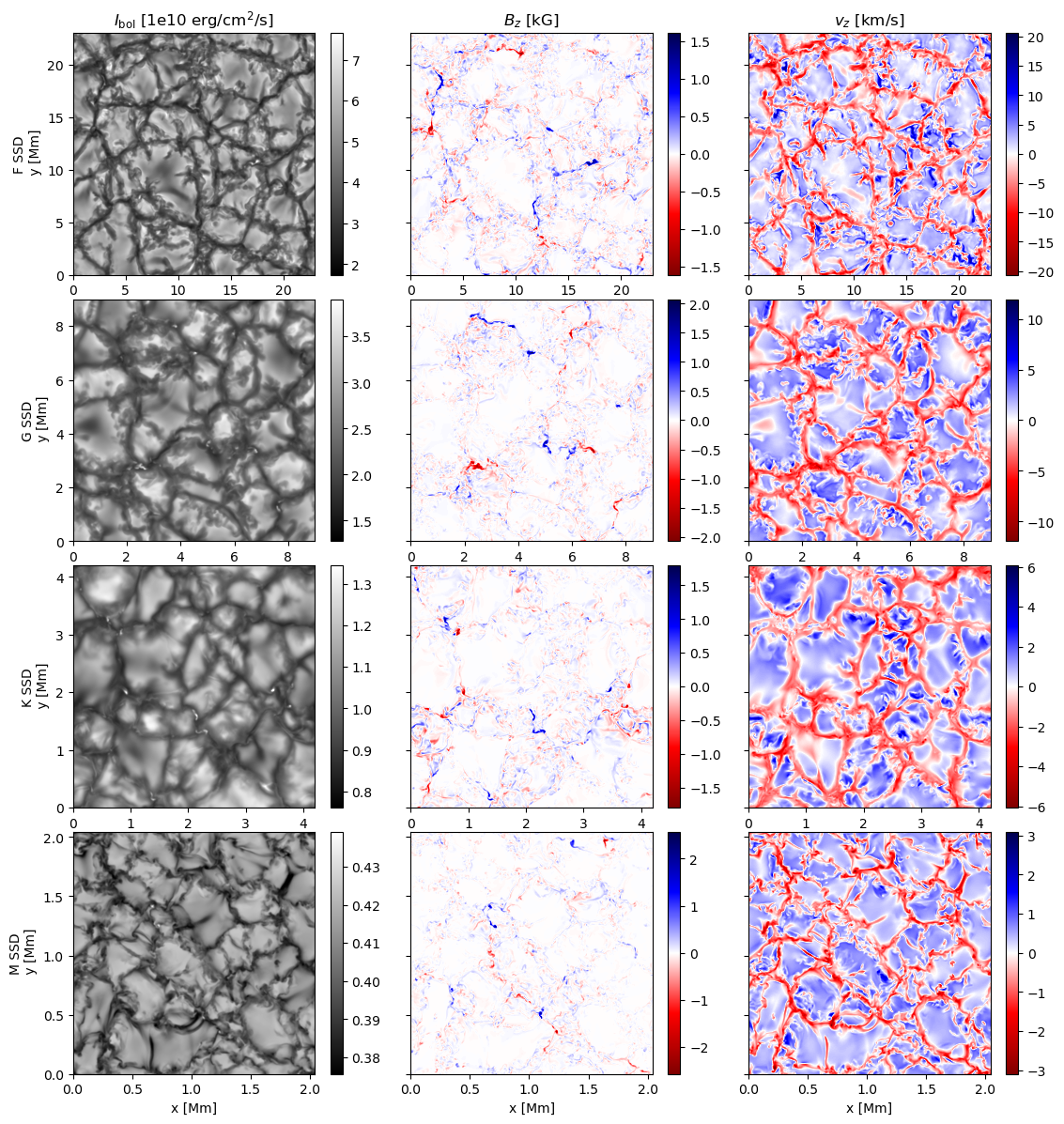}
    \caption{Same as Fig. \ref{fig:snap_300g}, but for the SSD cases.}
    \label{fig:snap_ssd}
\end{figure*}

\end{appendix}

\end{document}